\documentstyle[12pt]{article}

\begin{document}
\titlepage
\def\x{\times}
\def\ra{\rightarrow}
\def\beq{\begin{equation}}
\def\eeq{\end{equation}}
\def\beqa{\begin{eqnarray}}
\def\eeqa{\end{eqnarray}}
\def\D{ {\cal D}}
\def\L{ {\cal L}}
\def\C{ {\cal C}}
\def\calE{{\cal E}}
\def\lin{{\rm lin}}
\def\Tr{{\rm Tr}}
\def\mxth{\mathsurround=0pt }
\def\xversim#1#2{\lower2.pt\vbox{\baselineskip0pt \lineskip-.5pt
  \ialign{$\mxth#1\hfil##\hfil$\crcr#2\crcr\sim\crcr}}}
\def\simgr{\mathrel{\mathpalette\xversim >}}
\def\simle{\mathrel{\mathpalette\xversim <}}

\def\R{ {\cal R}}
\def\I{ {\cal I}}
\def\calR{ {\cal R}}
\def\lag{Lagrangian}
\def\Kahler{K\"{a}hler}


\def\a{\alpha}
\def\b{\beta}
\def\dota{ {\dot{\alpha}} }
\def\lag{Lagrangian}
\def\Kahler{K\"{a}hler}
\def\kahler{K\"{a}hler}
\def\A{ {\cal A}}
\def\C{ {\cal C}}
\def\D{ {\cal D}}
\def\F{{\cal F}}
\def\L{ {\cal L}}
\def\R{ {\cal R}}
\def\x{ \times }
\def\ra{\rightarrow}
\def\beq{\begin{equation}}
\def\eeq{\end{equation}}
\def\beqa{\begin{eqnarray}}
\def\eeqa{\end{eqnarray}}


\renewcommand{\thesection}{\arabic{section}.}
\renewcommand{\theequation}{\thesection \arabic{equation}}

\titlepage

\begin{flushright} CERN-TH.6961/93  \end{flushright}

\vspace{4ex}

\begin{center}

{\bf SUPERSYMMETRIC CALCULATION OF MIXED K\"{A}HLER-GAUGE
AND MIXED K\"{A}HLER-LORENTZ ANOMALIES} \\
         \rm
\vspace{3ex}
Gabriel Lopes Cardoso\footnote{Work supported
in part by the Department of Energy
under Contract No. DOE-AC02-76-ERO-3071. }\\ and \\
Burt A. Ovrut\footnote{On sabbatical leave from the Department of
Physics, University of Pennsylvania, Philadelphia, PA 19104-6396,
USA. }\\
Theory Division, CERN, CH-1211 Geneva 23, Switzerland \\

\vspace{3ex}

{\bf ABSTRACT}

\end{center}

We present a manifestly supersymmetric procedure for
calculating the contributions from matter loops to
the mixed K\"{a}hler-gauge and to the mixed K\"{a}hler-
Lorentz anomalies
in $N=1, D=4$ supergravity-matter systems.  We show
how this procedure leads to the well-known result for the
mixed K\"{a}hler-gauge anomaly.  For general supergravity-matter
systems the
mixed K\"{a}hler-Lorentz
anomaly is found to contain a term proportional to ${\cal R}^2$
with a background field
dependent coefficient as well as terms proportional to
$(C_{mnpq})^2$ and to the Gauss-Bonnet topological density.
We briefly comment
on the relationship between the mixed K\"{a}hler-Lorentz anomaly
and the moduli dependent threshold corrections to gravitational
couplings in $Z_N$ orbifolds.      \\
\vspace{25 mm}
\begin{flushleft} CERN-TH.6961/93 \\
August 1993 \end{flushleft}

\newpage

\section{Introduction}

\hspace*{.3in} As it is well known \cite{1,2},
the tree-level \lag\ describing the coupling of chiral and antichiral
matter supermultiplets $\Phi$ and ${\Phi}^{\dagger}$ to
$N=1, D=4$ supergravity is specified
by three fundamental functions, namely the K\"{a}hler potential $K$,
the holomorphic superpotential $W$ and the holomorphic
gauge coupling function $f$.  In the conventional formulation
of the theory, the tree-level \lag\ is invariant under combined
super-Weyl-K\"{a}hler transformations \cite{1,2}.  These consist of
superfield rescalings of the supervielbein, which leave the
conventional torsion constraints invariant, followed by
K\"{a}hler transformations of the three fundamental
functions $K,W,f$.  The chiral matter superfields do not
transform under either
the super-Weyl or the
K\"{a}hler transformations.
On the other hand, the tree-level \lag\
in the conventional formulation has, in general, a
non-canonical gravitational kinetic energy term of the form
$h(A,\bar A) {\cal R}$, where $A$ represents component
matter scalar
fields, $h$ is some calculable function and $\cal R$ is the
curvature scalar.
It is also well known \cite{1,2}
that, by appropriate rescaling of the graviton and of
the gravitino
as well as of the auxiliary fields $M$ and $b_a$ of
minimal supergravity,
one can obtain a new \lag\ with appropriately Einstein
normalized gravitational kinetic energy, $\R$.
This rescaling can either be performed at the component
field level \cite{1,2}
or at the superfield level \cite{4,5,6}, the latter
having the advantage of being manifestly supersymmetric.
It turns out that
this rescaled \lag\ is invariant under an important new symmetry,
called K\"{a}hler symmetry \cite{1,2,3}.  By
explicitly adding the K\"{a}hler transformations to the structure
group of superspace \cite{4,5,6},
one can give a complete superfield description of this
appropriately Einstein normalized supergravity-matter theory.
The new underlying superspace is called K\"{a}hler superspace.
The K\"{a}hler invariance
of the Einstein normalized tree-level \lag\ can be traced back
to the super-Weyl-K\"{a}hler invariance of the tree-level
\lag\ in the conventional superspace formulation of the theory
\cite{1,2,90}.
Hence, a chiral matter superfield in K\"{a}hler superspace
carries a zero K\"{a}hler charge.  Its fermionic component
field, however, carries a non-vanishing K\"{a}hler charge and,
 consequently, couples to the component
K\"{a}hler connection $a_a$, which is
a composite connection made out of component matter fields.
Other fermions, such as gauginos and the gravitino, also carry
a non-vanishing K\"{a}hler charge and, hence, also couple to the
K\"{a}hler connection $a_a$.
   As always in four dimensions, the K\"{a}hler symmetry of the
Einstein normalized
tree-level  \lag\
might get spoiled by one-loop component
triangle graphs with fermions running in the loop and with
at least one
K\"{a}hler connection sticking out.
This is, in fact, what happens in general supergravity-matter
theories \cite{7,8,9,10}.  On the other hand,
since only fermions carry K\"{a}hler charge,
it is impossible to construct component graphs having legs of the
K\"{a}hler type sticking out and having component fields running
in the loop other than fermions.  This, however, creates a
puzzle.  Namely, it would appear that it is then
impossible to construct a supersymmetric expression for the
component K\"{a}hler anomaly in
K\"{a}hler superspace because of missing component graphs.

In this paper, we will present a manifestly supersymmetric
procedure for calculating
mixed K\"{a}hler-gauge and mixed K\"{a}hler-Lorentz anomalies.
These are the ones
generated by component fermionic triangle graphs with only
one K\"{a}hler connection sticking out.  These mixed K\"{a}hler
anomalies have important
physical consequences.  They have been shown to lead to threshold
corrections to gauge and gravitational couplings
\cite{7,8,9,10,11,12,13,40} in
string theory
as well as to inflationary supergravity models \cite{14,15}.  We will
restrict
our discussion to the contribution from matter loops to these
mixed anomalies.  We believe, though, that the procedure
described in this paper can also be extended to include contributions
from gauge and gravitational fields running in the loop.

The origin of the puzzle is not hard to understand.  As mentioned
above, a chiral matter superfield carries vanishing K\"{a}hler
charge $\omega$
in K\"{a}hler superspace.  This means that there is no
direct coupling ${\Phi}^{\dagger} e^{{\omega} K} \Phi$
of the K\"{a}hler prepotential $K$ to
quadratic matter superfields, since $\omega (\Phi) = 0$.
This is to be contrasted with the coupling of the
gauge prepotential $V$ to charged
matter as ${\Phi}^{\dagger} e^V \Phi$.
Were one to examine the pure gauge anomaly three-point function,
one would find a supersymmetric result, since it could
have been obtained by
calculating a superfield triangle
graph with $\Phi$ running in the loop and $V$-legs sticking
out.  It seems then clear that, in order to directly calculate
supersymmetric expressions for the mixed K\"{a}hler anomalies,
one must go to a formalism where $K$ couples to chiral matter
as ${\Phi}^{\dagger} e^K \Phi$.  The conventional superspace formalism
is such a framework.  It allows for a superfield prepotential
formalism describing the coupling of
quantum matter superfields $\phi$ to
background prepotentials such as $K$.  An example of such a coupling
is precisely ${\phi}^{\dagger}  e^{-\frac{1}{3} K} \phi$.
Since there exists a superfield prepotential formalism,
all calculations of anomalous triangle graphs in the conventional
formulation can, in
principle, be performed at the superfield level, yielding
automatically supersymmetric results.

The relevant tree-level symmetry
in the conventional superspace formulation is
the tree-level super-Weyl-K\"{a}hler symmetry.  Hence,
of relevance
in the conventional formulation are the couplings of
quantum matter superfields to
K\"{a}hler as well as supergravity
prepotentials, and one has to
look out for the
breakdown of the tree-level super-Weyl-K\"{a}hler symmetry
due to the non-vanishing of mixed triangle graphs with
quantum matter superfields running in the loop.
We do indeed find that such mixed triangle graphs are non-vanishing,
and, hence, that
they contribute to the breakdown of the super-Weyl-K\"{a}hler
symmetry.  As stated above, results in the conventional
superspace formulation of the theory can be transformed over
into K\"{a}hler superspace by particular superfield rescalings
of the underlying torsion constraints
of conventional superspace \cite{4,5,6}.
Upon applying these rescalings to the mixed super-Weyl-K\"{a}hler
anomalies in conventional superspace, we arrive at manifestly
supersymmetric results in K\"{a}hler superspace which qualify
to be called the supersymmetric mixed K\"{a}hler-gauge
and the supersymmetric mixed K\"{a}hler-Lorentz anomaly,
respectively.  In this way, we give the first complete
calculation for, and rediscover the well-known
result \cite{7,9,10}
for the supersymmetric
mixed K\"{a}hler-gauge anomaly.  We also give the first
complete calculation for the supersymmetric mixed
K\"{a}hler-Lorentz anomaly.
When putting  the gravitational fields on-shell, the
supersymmetric mixed K\"{a}hler-Lorentz anomaly
calculated using our procedure
also reduces to its
well-known form given in \cite{8,9}.  Off-shell, however,
it contains an additional term proportional to ${\cal R}^2$
which is not part of the Gauss-Bonnet topological density.
Two remarks concerning this term need to be made.
First of all,
this ${\cal R}^2$-term, present
in general supergravity-matter theories,
arises from graphs constructed out of
vertices of the type $E e^{-\frac{1}{3} K}{\phi}^2 $
in the conventional formulation.  Such vertices are not invariant
under conformal transformations (these are defined in section 6),
and, hence, the growing of
a term proportional to ${\cal R}^2$ not contained in the
Gauss-Bonnet combination is not forbidden by any
symmetry argument.  Secondly, the
coefficient of this ${\cal R}^2$ term comes out to be
background field dependent.

We close this paper with a remark on the relationship
between the supersymmetric mixed K\"{a}hler-Lorentz anomaly,
as computed in this paper, and the moduli dependent threshold
corrections to gravitational couplings in $Z_N$ orbifolds
computed in \cite{13}.  These threshold corrections turn out to be
proportional to the trace anomaly coefficients of the different
fields coupled to
gravity.  From the field theory analysis presented in this
paper this
can be understood by noticing that there are two
distinct terms which contribute to the K\"{a}hler anomaly.
  One of them comes with a coefficient
proportional to the trace anomaly whereas the other one
comes with a coefficient proportional to the chiral anomaly
in conventional superspace.  It is this latter contribution
which gets removed by the Green-Schwarz mechanism \cite{7,8,9},
yielding threshold corrections proportional to the trace anomaly
coefficients of the various fields coupled to gravity.

\section{$U_K(1)$ Superspace and Tree-Level Symmetries}

\hspace*{.3in} In this section, we will briefly review
some of the features of K\"{a}hler superspace geometry which will
be relevant in the subsequent discussion.  A complete description
of the properties of $U_K(1)$ superspace can be found in
\cite{6}.  The structure group of \Kahler\ superspace is
taken to be $SL(2,C) \x U_K (1)$ and, accordingly, one
introduces two Lie algebra valued one-form gauge connections
$\phi_B \,^A = dz^M \phi_{MB}\,^A$ and $A = dz^M$ $A_M$
corresponding to the Lorentz and $U_K (1)$ groups, respectively.
In addition, one introduces a supervielbien $E_M \,^A$ and the
associated one-forms $E^A = dz^M E_M \,^A$.  The $U_K (1)$ gauge
connection $A$ is a composite gauge connection defined by
\beqa A_\alpha &=& \frac{1}{4} \D_\alpha K \nonumber \\
 A^{\dot{\alpha}} &=& -\frac{1}{4} \bar{\D}^{\dot{\alpha}} K
\nonumber \\
A_{\alpha \dot{\alpha}} &=& - \frac{i}{8} \left[ \D_{\alpha},
\bar{\D}_{\dot{\alpha}} \right] K   \eeqa   
where the prepotential $K(\Phi_i, \; {\Phi_i}^{\dagger})$
is the \Kahler\
potential for matter chiral superfields $\Phi_i$.  All matter
superfields have vanishing $U_K (1)$ weight, $\omega_K(\Phi_i) =
0$.  Under a \Kahler\ transformation
\beq {\kappa}^2 K(\Phi_i,\; {\Phi_i}^{\dagger})
\ra {\kappa}^2 K (\Phi_i, \; {\Phi_i}^{\dagger}) +
F(\Phi_i) + \bar{F} ({\Phi_i}^{\dagger}) \eeq 
the one-form $A$ transforms as
\beq A \ra A + {\kappa}^{-2} \frac{i}{2} d  \; Im \, F \eeq 
where $Im \, F = \frac{F- \bar{F}}{2i}$ and
${\kappa}^2 = 8 \pi M_P^{-2}$.  $M_P$ is the Planck mass.
Also, under a \Kahler\
transformation the supervielbein one-forms $E^A$ can be shown
\cite{6} to transform as
\beq E^A \ra E^A \exp \left[ - \frac{i}{2} \omega (E^A) Im\, F
\right] \eeq 
where
\beq \omega (E^\alpha) = 1, \; \; \omega (E_{\dot{\alpha}}) = -1
\;\; \omega (E^a) = 0\;\; . \eeq 
Solving the Bianchi identities subject to a set of constraints
\cite{60}, one finds that all components of the torsion and
curvature may be expressed in terms of a set of superfields and
their coordinate derivatives:
\beq \begin{array}{lcccccc}
{\rm superfield} & \;\;\; & R & R^{\dagger} & G_{\alpha \dot{\alpha}} &
W_{\alpha \beta \gamma} \; , \; X_\alpha & \bar{W}_{\dot{\alpha}
\dot{\beta} \dot{\gamma}}\; , \; \bar{X}_{\dot{\alpha}} \\
U_K(1) \; {\rm weight} & & 2 & -2 & 0 & 1 & -1 \end{array} \eeq
where
\beqa X_\alpha &=& \D_\alpha \R - \bar{\D}^{\dot{\alpha}}
G_{\alpha \dot{\alpha}} = - \frac{\kappa^2}{8}\left( \bar{\D}^2 -
8R \right) \D_\alpha K \nonumber \\
\bar{X}^{\dot{\alpha}} &=& \bar{\D}^{\dot{\alpha}} R^{\dagger} +
\D_\alpha G^{\alpha \dot{\alpha}} = - \frac{\kappa^2}{8}
\left(\D^2 - 8R^{\dagger}\right) \bar{\D}^{\dot{\alpha}} K \eeqa 
$X_{\alpha}$ is the superfield fieldstrength of the $U_K(1)$
gauge connection.  If we further assume that there is an internal
gauge group, then we must introduce yet another Lie algebra
valued one-form gauge connection $\A_a\,^b = dz^M \A_{M
a}\,^b$.  Solving the Bianchi identities now introduces a new
superfield fieldstrength, $W^a_\alpha$, with $U_K(1)$ weight
$\omega (W_\alpha^a) = 1$.

Using these superfields, one can write down the tree-level
superfield \lag\ in K\"{a}hler superspace
\cite{4,5,6}.  It consists of three
parts, each specified by a fundamental and independent function.
The first part, specified entirely by the K\"{a}hler potential, is
the supergravity-matter kinetic energy term given by
\beq \L_0 = -3 \kappa^{-2} \int d^4 \theta E \left[K\right] \eeq
where $E$ is the superdeterminant.  The second part, specified by
the holomorphic superpotential $W(\Phi_i)$, is the potential
energy term given by
\beq \L_{PE} = \frac{1}{2} \int d^4 \theta \frac{E}{R}
e^{{\kappa}^2 \frac{K}{2}} W(\Phi_i) + h.c. \eeq 
Finally, the Yang-Mills \lag\ is given by
\beq \L_{YM} = \frac{1}{8} \int d^4 \theta \frac{E}{R}
f(\Phi_i)_{ab} \; W^{\alpha a} W_\alpha\,^b + h.c. \eeq 
where $f(\Phi_i)_{ab}$ is the holomorphic gauge coupling
function.  The total \lag\ possesses, at the tree-level, three
symmetries.

\vspace{4ex}

\noindent 1. \ \ \Kahler\ invariance:

\vspace{4ex}

\noindent Under \Kahler\  transformation
${\kappa}^2 K \ra {\kappa}^2 K + F + \bar{F}$,
it can be shown \cite{4,5,6} that
\beqa E & \ra & E \nonumber \\
R & \ra & Re^{-(F-\bar{F})/2} \eeqa 
One also, simultaneously, transforms $W$ and $W_\alpha$ as
\beqa W & \ra & e^{-F} W \nonumber \\
W_\alpha & \ra & e^{-(F-\bar{F})/4} W_\alpha \; .\eeqa 
Then
\beq \L'_0 = - 3 \kappa^{-2} \int d^4 \theta E' = -3 \kappa^{-2}
\int d^4 \theta E = \L_0 \eeq 
and
\beqa \L'_{PE} &=& \frac{1}{2} \int d^4 \theta \frac{E'}{R'}
e^{{\kappa}^2 K'/2} W' + h.c. \nonumber \\
&=& \frac{1}{2} \int d^4 \theta \frac{E e^{{\kappa}^2 {K/2}}  W \;
e^{(F+\bar{F})/2} e^{-F} } {R \; e^{-(F- \bar{F})/2}}+ h.c. =
\L_{PE} \eeqa 
Furthermore
\beqa \L'_{YM} &=& \frac{1}{8} \int d^4 \theta \frac{E'}{R'}
f'W'^{\alpha} \; W'_{\alpha} + h.c. \nonumber \\
&=& \frac{1}{8} \int d^4 \theta \frac{EfW^\alpha W_\alpha e^{-(
F-\bar{F})/2} }{ Re^{-(F - \bar{F}) /2}}+ h.c. = \L_{YM} \eeqa
That is, the complete superfield \lag\ is invariant under the
\Kahler\ transformations
${\kappa}^2 K \ra {\kappa}^2 K + F + \bar{F}$ and (2.12).

\vspace{4ex}

\noindent 2. \ Gauge invariance:

\vspace{4ex}

\noindent This follows from the fact that both $E$ and $R$ are
functions of the \Kahler\ potential $K$ which, in turn, is
invariant under Yang-Mills transformations of the charged
superfields $\Phi_k$.  These are given by
\beqa e^V & \ra & e^{-i \bar{\Lambda}} e^V e^{i \Lambda}
\nonumber \\ \Phi_k & \ra& e^{-i \Lambda} \Phi_k \eeqa 
where
\beq \bar{\D}^{\dot{\alpha}} \Lambda = 0 \eeq 
and $V = V^{(r)} T^{(r)}$ denotes the Yang-Mills prepotential
vector superfield (the $T^{(r)}$ denote the hermitian generators
of the Yang-Mills gauge group $H$).  The square of the Yang-Mills
superfield fieldstrength, $W^\alpha W_\alpha$, is by construction
invariant under Yang-Mills transformations.

\vspace{4ex}

\noindent 3. \ Lorentz invariance:

\vspace{4ex}

\noindent This follows from the fact that all the superfield
combinations appearing in $\L_0$ and $\L_{YM}$ as well as in
$\L_{PE}$ are scalars with respect to Lorentz transformations.

Invariances (1) -- (3) of the tree-level theory of the
supergravity-matter supermultiplet system can, of course, also be
displayed at component level.  Component fields are defined
according to standard notation \cite{4,5,6}:  $A^i$,
$\chi^i_\alpha$, $\F^i$ for chiral multiplets (and similar
notations for antichiral multiplets) and $\lambda^\alpha$, $v_m$,
$\D$ for Yang-Mills multiplets.  The irreducible minimal
supergravity multiplet is realized by $(e_m\,^a, \;
\psi^\alpha_m, \; M, \; b_a)$.  $M$ and $b_a$ denote the
auxiliary component fields of minimal supergravity.  The
covariant derivative of a generic Weyl fermion will, in a theory
with invariances (1) -- (3), then contain a connection for each
of these symmetries.

\vspace{4ex}

\noindent 1. \ The component connection for gauging \Kahler\
transformations is given by the lowest component
\cite{4,5,6} of the $U_K(1)$ gauge connection superfield
${A}_{\alpha\dot{\alpha}}$
\beq \left. A_{\alpha \dot{\alpha}} \right| = - \frac{i}{8}
\left[ \D_\alpha, \bar{\D}_{\dot{\alpha}} \right] \left. K
\right| = a_{\alpha \dota} = \sigma^m_{\alpha \dota} a_m \eeq
where
\beq a_m = \frac{1}{4} \left( \partial_j K \D_m A^j -
\partial_{\bar{j}} K \D_m \bar{A}^{\bar{j}} \right) + i
\frac{1}{4} g_{i \bar{j}} \left(\chi^i \sigma^m
\bar{\chi}^{\bar{j}} \right) \eeq 
Here, $g_{i\bar{j}}$ denotes the \Kahler\ metric $g_{i \bar{j}}
= \partial_i \partial_{\bar{j}} K$ of the matter manifold
parameterized by $A^i$ and $\bar{A}^{\bar{j}}$.  Under \Kahler\
transformations
\beq a_m\, ' = a_m + {\kappa}^{-2}
\frac{i}{2} \partial_m \; Im \, F.\eeq 
All component Weyl fermions transform under \Kahler\
transformations (2.2).  The component matter Weyl fermion,
$\chi_\alpha^i = (\frac{1}{\sqrt{2}}) \D_\alpha \Phi_i|$,
transforms as
\beq \chi'^i = e^{(i/2 \; Im \, F)} \chi^i \eeq 
whereas the gaugino $\lambda^\alpha$ transforms with opposite
charge
\beq \lambda'^\alpha = e^{(- i/2 \; Im \, F)} \lambda^\alpha \eeq
as does the gravitino $\psi^\alpha_m$.

\vspace{4ex}

\noindent 2. \ The component Yang-Mills connection $v_{\alpha
\dota} = v^{(r)}_{\alpha \dota} T^{(r)}$ is contained in the
Yang-Mills prepotential $V$ as
\beq \frac{1}{2} \left[ \D_\alpha, \bar{\D}_\dota \right] \left.
V \right| = - 2 v_{\alpha \dota} \eeq 

\vspace{4ex}

\noindent 3. \ The component Lorentz connection $\omega_{m
\alpha} \,^\beta$ is given by the lowest component of the
$SL(2,C)$ gauge connection superfield $\phi_{m \alpha}\,
 ^\beta$
\beq \left. \phi_{m \alpha} \,^\beta \right| = \omega_{m \a}
\,^\b  \eeq 
Therefore, the covariant derivative for matter fermions
$\chi^i_\alpha$ reads \cite{4,5,6}
\beqa \D_m \chi^i & = & \left( \partial_m \right.
+ iv_m^{(r)} \left( T^{(r)} - \frac{1}{2} D^{(r)}\right)
\nonumber \\ &-& \left. \omega_m + \frac{i}{2} b_m - \kappa^2
a_m \ldots \right) \chi^i \eeqa 
where
\beq D^{(r)} = \kappa^2 \frac{\partial K}{\partial A^{ia} }
T^{(r)a} \,_b \, A^{ib} \eeq 
The $\ldots$ stand for the additional coupling to the
$\sigma$-model
Christoffel connection $\Gamma^i_{jk} = g^{i \bar{j}}
\partial_{\bar{j}} g_{jk}$.  Such a coupling will not be
considered in this paper.

The covariant derivative of the matter scalar $A^i$ is given by
\beq \D_m A^i = \partial_m A^i + iv^{(r)}_m T^{(r)} A^i \eeq
Note the absence of a coupling of the matter scalar $A^i$ to the
\Kahler\ connection $a_m$ given in (2.19).  This is in manifest
contrast to the matter fermion $\chi^i$.  The reason for this is
that the \Kahler\ charge $\omega_K (\Phi_i)$ of matter superfield
$\Phi_i$ is zero, $\omega_K (\Phi_i) = 0$.  Therefore, only the
component fermions carry \Kahler\ charge and, hence, only
component fermions rotate under \Kahler\ transformations (2.2).

We will now expand the supergravity-matter kinetic superfield
\lag\ (2.8) into components.  It is, by now, well known that one
of the advantages of the \Kahler\ superspace formulation lies in
the fact that it immediately gives the correctly normalized
kinetic terms for all the component fields without any need for
rescalings or complicated partial integrations at the component
field level.  That is,
it immediately gives the correctly normalized Einstein-Hilbert
action as well as making the component \Kahler\ structure in the
matter sector manifest.  The component kinetic \lag\ for the
supergravity-matter system reads \cite{1,2,3,4,5,6}
\beqa \L_0/e &=& - \frac{1}{2} \kappa^{-2} e \R - \frac{1}{3}
\kappa^{-2} \left( M \bar{M} - b^a b_a \right) \nonumber \\
&-& g^{mn} g_{i \bar{j}} \D_n A^i \D_m \bar{A}^{\bar{j}} -
\frac{i}{2} \chi^{\alpha i} g_{i\bar{j}} \sigma^m_{\alpha \dota}
\D_m \bar{\chi}^{\bar{j} \dota} + \frac{i}{2} \left( \D_m
\chi^{\alpha i} \right) g_{i \bar{j}} \sigma^m_{\alpha \dota}
\bar{\chi}^{\bar{j} \dota} \nonumber \\
&+& {\cal D}^{(r)} D_{(r)} + \ldots \eeqa
where we have only displayed the component terms relevant for
this paper.

Of importance to this paper are the couplings of matter currents
to external gravitational fields and to Yang-Mills and \Kahler\
connections.  Let us look at the fermionic matter current $g_{i
\bar{j}} \chi^i_\a \bar{\chi}^{\bar{j}}_\dota$.  From the
component \lag\ (2.28)
it follows that the fermionic current couples to
the Yang-Mills connection $v_m^{(r)}$, the \Kahler\ connection
$a_m$, to the gravitational spin connection $\omega_{m \a} \,^\b$
as well as to the auxiliary field $b_a$ and the space-time
metric.  The bosonic matter
current $g_{i \bar{j}} A^i \partial_m A^{\bar{j}}$, on the other
hand, couples only
to the Yang-Mills connection $v^{(r)}_m$, as well
as to the space-time metric $g_{mn}$.

Let us emphasize again how differently both component currents
couple to the \Kahler\ connection.  The bosonic component current
doesn't couple to $a_m$ at all!  This is in manifest contrast to
the coupling of both currents to the Yang-Mills connection $v_m$.
Also note that there is a coupling of $A^i \bar{A}^{\bar{j}}$ to
the $\D$-term of the Yang-Mills prepotential
$V$, through the term $\D^{(r)} D_{(r)}$, whereas there is no
such coupling of $A^i \bar{A}^{\bar{j}}$ to the highest component
of prepotential $K$.  The reason for all of this is, once again,
obvious.  On one hand, the matter superfield $\Phi_i$ carries a
non-zero Yang-Mills charge, on the other hand it has zero
\Kahler\ charge, $\omega_K (\Phi_i) = 0$.  This will be of utmost
importance in the next section.

Finally, note that we have not expanded
the potential energy (2.9)
into component fields, since its component expansion will not be
needed in this paper.  Its superfield form is, however, relevant
to our discussion.
The additional matter couplings contained
in the Yang-Mills Lagrangian (2.10), on the other hand, are
of no relevance for the subsequent discussion of the
K\"{a}hler anomalous contributions
from triangle graphs to the one-particle irreducible
effective action.  Hence,
we will ignore them throughout this paper.

\setcounter{equation}{0}

\section{\Kahler\ Anomalies in the \Kahler\ Superspace Formalism}

\hspace*{.3in}
Since all of the fermions in the supergravity-matter theory are
chiral, the possibility exists that all three of the symmetries
discussed in the previous section are anomalous at the one-loop
level.  However, it is well known that there are no pure Lorentz
anomalies in four-dimensions.  Furthermore, we will assume, as in
the standard electroweak model, that the matter superfield content
is so chosen that there are no pure gauge anomalies or mixed
gauge-Lorentz anomalies.  What about possible one-loop anomalies
in the \Kahler\ symmetry?  It has recently been demonstrated
\cite{7,8,9,10} that
both non-vanishing pure \Kahler, mixed \Kahler-gauge and
\Kahler-Lorentz anomalies, in general, exist.  We will not
discuss the pure \Kahler\ anomalies in this paper, since their
physical relevance is presently
obscure.  The mixed \Kahler\ anomalies,
however, are known to lead to important phenomenological and
cosmological effects
\cite{7,9,10,14,15,40},  and it is these mixed anomalies we now
analyze in detail.

We begin by considering the mixed \Kahler-gauge anomaly.
We will, throughout this paper, work to lowest
order in supergravity and in K\"{a}hler background fields.
Consider a set of matter chiral superfields, $\Phi_i$, a subset
of which have
non-vanishing gauge charges, so chosen as to satisfy the
condition that they are free of pure gauge anomalies.
The relevant
part of the supergravity-matter \lag\ is given in (2.28), with
the covariant derivatives given by (2.25) and (2.27).
The graph
in Figure~1a gives the fermionic contribution to the mixed
\Kahler-gauge anomaly.  This graph can be evaluated \cite{17}
and the
contribution to the
associated effective \lag\
is found to be
\beq \L_\chi = \frac{i \kappa^2}{(4\pi)^2} \; Tr \; F_{mn}
\tilde{F}^{mn} \frac{1}{\Box_0} \partial^p a_p \eeq 
where $F_{mn}$ is the covariant curl of $v_m$,
$\tilde{F}^{mn}$ $= \frac{1}{2} \epsilon^{mn\ell p} F_{\ell p}$ and
$\Box_0$ denotes the flat space d'Alembertian.  Implied in (3.1)
is a sum over all charged
chiral matter multiplets $\Phi_i$.
It can readily be
seen that (3.1) is anomalous under \Kahler\ transformations.
Using (2.20) we find that
\beq \delta_K \L_\chi = - \frac{1}{2} \;
\frac{\kappa^2}{(4\pi)^2} \; Im \, F \; Tr \; F_{mn}
\tilde{F}^{mn} \eeq 
which, since it doesn't vanish, implies that \Kahler\
symmetry is broken.  Now $\L_\chi$ by itself is not
supersymmetric.  Before trying to find the one-loop graphs that
will supersymmetrize (3.1), it is worth noting that there is a
unique superfield expression whose highest component contains
$\L_\chi$.  This expression is given by
\beq \L = - \frac{1}{8} \frac{\kappa^2}{(4\pi)^2} \int d^4 \theta
\, Tr \, W_\a^2 \frac{1}{\Box_0} D^2K + h.c. \eeq 
Expanding (3.3) out into component fields yields
\beqa \L &=& \frac{i \kappa^2}{(4\pi)^2} \, Tr \, F_{mn} \,
\tilde{F}^{mn} \frac{1}{\Box_0} \partial^p a_p \nonumber \\
&-& \frac{\kappa^2}{8(4\pi)^2} \, Tr \, F_{mn} \, F^{mn}
\frac{1}{\Box_0} \left( 4 K \left|_{\theta^2 \bar{\theta}^2} +
\Box_0 K \right| \right) \nonumber \\
&+& \ldots \eeqa 
where $\ldots$ refers to terms containing fermions.  $K|$and
$K|_{\theta^2\bar{\theta}^2}$ are the lowest and $\theta^2
\bar{\theta}^2$ components of superfield $K (\Phi_i,
\Phi^{\dagger}_i)$ respectively given by
\beqa \left. K \right| &=& K(A^i, \bar{A}^{\bar{i}} ) \nonumber
\\
\left. K \right|_{\theta^2 \bar{\theta}^2} &=& - g_{i\bar{j}}
\D_m A^i \D^m \bar{A}^{\bar{j}} + g_{i \bar{j}} {\cal F}^i
\bar{{\cal F}}^{\bar{j}} + \frac{1}{4} \Box_0 K + \ldots \eeqa 
where $\ldots$ stand for terms containing fermions.  Now the term
in (3.4) proportional to $Tr$ $F_{mn} F^{mn}$ is $CP$ even and
would have to be generated by a one-loop graph with scalar fields
running around the loop.  But, it is clear from (2.28) that there
is no quadratic coupling of $A^i$ to either $K|$ or $K|_{\theta^2
\bar{\theta}^2}$.  Therefore, there are no one-loop graphs that
can generate the second term in (3.4)!  It is, apparently,
impossible to construct a supersymmetric expression for the mixed
\Kahler-gauge anomaly using one-loop graphs generated from the
K\"{a}hler superspace \lag.  There have been some attempts to show
that the missing graphs arise from gauge field two-point
functions through the regularization procedure.  However, these
attempts explicitly break gauge invariance when $K(A^i,
\bar{A}^{\bar{i}})$ is non-constant and must, in our opinion, be
discarded.  Before showing how to perform a supersymmetric
calculation, we demonstrate that the same problem exists for the
mixed \Kahler-Lorentz anomalies.

Again, the relevant part of the supergravity-matter \lag\ is
given in (2.28) with the covariant derivatives given by (2.25)
and (2.27).  The graph in Figure~1b gives the fermionic
contribution to the mixed \Kahler-Lorentz anomaly.
This graph can be evaluated \cite{17} and the
contribution to the
associated effective \lag\ , steming from one
Weyl fermion running in the loop,
is found to be
\beq \L'_\chi = - i \frac{\kappa^2}{24(4\pi)^2} \R_{mn \; b}\,^a \;
\tilde{\R}^{mn}\,_a\,^b \frac{1}{\Box_0} \partial^p a_p \eeq 
where $\R_{mnb}\,^a$ is the Lie algebra valued curvature tensor
and $\tilde{\R}^{mn}\,_a\,^b = \frac{1}{2} \epsilon^{mn\ell p} \R_{\ell
pa} \,^b$.  If there are $N$ chiral matter fermions in the theory,
then (3.6) would simply be multiplied by $N$.
It is easy to see using (2.20) that (3.6) is not
invariant under \Kahler\ transformations.  Now, $\L'_\chi$ by itself is
not supersymmetric.  Before trying to find the one-loop graphs
that will supersymmetrize (3.6), it is worth noting that there is
a minimal superfield expression whose highest component contains
$\L'_\chi$.  This expression is given by
\beq \L' = - \frac{\kappa^2}{24(4\pi)^2} \int d^4\theta \;
W_{\a\b\gamma}^2 \frac{1}{\Box}_0 D^2 K + h.c. \eeq 
Expanding (3.7) out into component  fields \cite{25,70}
yields
\beqa \L' &=& - \frac{i \kappa^2}{24(4\pi)^2} \R_{mnb}\,^a
\tilde{\R}^{mn}\,_a\,^b \frac{1}{\Box_0} \partial^p a_p \nonumber
\\
&+& \frac{\kappa^2}{192(4\pi)^2} C_{mnpq} C^{mnpq}
\frac{1}{\Box_0} \left( 4 \left. K \right|_{\theta^2
\bar{\theta}^2} + \Box_0 K| \right) \nonumber \\
&+& \ldots \eeqa 
where $\ldots$ refers to terms containing fermions and
$\C_{mnpq}$ denotes the Weyl tensor.  Now the term in (3.8)
proportional to $C_{mnpq} C^{mnpq}$ is $CP$ even and would have to
be generated by a one-loop graph with scalar fields running
around the loop.  But,again, it is clear from (2.28) that there is no
quadratic coupling of $A^i$ to either $K|$ or $K|_{\theta^2
\bar{\theta}^2}$.  Therefore, there are no one-loop graphs that
can generate the second term in (3.8)!  Again, it is impossible
to construct a supersymmetric expression for the mixed
\Kahler-Lorentz anomaly using one-loop graphs generated from the
\Kahler\ superspace \lag.  Furthermore, any attempt to generate
the missing graphs from the graviton two-point function using the
regularization procedure will break Lorentz invariance.

As paradoxical as these results might seem, it is not too hard to
understand their origin.  In superfields, the gauge prepotential
couples to matter as $\Phi^{\dagger} e^V \Phi$.  Were we to examine
the pure gauge anomaly three-point function, we would to lowest
order in $V$ get as a result the supersymmetric expression
\beq \L'' \sim \int d^4 \theta \, Tr \, W^2_\a \frac{1}{\Box} D^2
V + h.c. \eeq 
If the \Kahler\ prepotential also coupled to matter as
$\Phi^{\dagger} e^K \Phi$, then we would indeed have arrived at the
supersymmetric expressions (3.3) and (3.7) for the mixed
\Kahler-gauge and \Kahler-Lorentz anomalies.  However, in
\Kahler\ superspace $\omega_K (\Phi_i) = 0$!  It follows that there
is no such direct coupling of $K$ to quadratic matter superfields
and, hence, one expects difficulty computing a supersymmetric
expression for the anomaly.  Two things, then, seem clear.
First, it must be possible to directly calculate a
supersymmetries expression for the mixed \Kahler\ anomalies.
Secondly, it seems likely that we must go to a formalism where
$K$ couples to matter as $\Phi^{\dagger} e^K \Phi$.  We show how to
do this in the next section.

\section{Conventional Superspace and Tree-Level Symmetries}

\setcounter{equation}{0}

\hspace*{.3in} As discussed in the introduction,
the \lag\ describing the coupling of $D
=4$, $N=1$ supergravity to matter supermultiplets has, in
general, a non-canonical gravitational kinetic energy of the form
$h(A,{\bar A})\R$.  In
the previous sections we showed that if we
worked with appropriately rescaled fields so that
$h(A,{\bar A}) =1$, then
the tree-level \lag\ was explicitly invariant under \Kahler\
symmetry.  However, computing the one-loop supersymmetric mixed
\Kahler\ anomalies with this \lag\ is apparently impossible, for
the reasons discussed above.  In this section, we go back to the
generic supergravity-matter \lag\ where
$h(A,{\bar A}) \neq 1$ and gravity
is not canonically normalized.  We will show that, in this case,
the \Kahler\ potential couples to quantum
matter in the generic form
$\Phi^{\dagger} e^K \Phi$ and that accordingly pure \Kahler\
symmetry is replaced by a mixed super-Weyl-\Kahler\ symmetry at
the tree-level.  This, as we will see, is the first step toward a
coherent calculation of mixed supersymmetric \Kahler\ anomalies.

In the generic case, the structure group of superspace is taken
to be simply $SL(2,C)$ with the associated one-form gauge
connection $\phi_B\,^A = dz^M \phi_{MB}\,^A$.  In addition, one
introduces the supervielbein $E_M\,^A$ and the associated
one-forms $E^A = dz^M E_M\,^A$.  Solving the Bianchi identities
subject to a set of constraints \cite{16}, one finds that
all components of the torsion and curvature may be expressed in
terms of a set of superfields and their coordinate derivatives:
\beq {\rm superfield} \;\;\; R,R^{\dagger} \;\;\; G_{\a\dota} \;\;\;
W_{\a\b\gamma}, \;\; {\bar{W}}_{\dota \dot{\beta} \dot{\gamma}}
\eeq 
Since they are obtained by solving the Bianchi identities with
respect to a very different set of constraints, the $R$, $G_{\a
\dota}$ and $W_{\a \b \gamma}$ in this section are different
than, and not to be confused with, the field strength solutions
of the Bianchi identities in \Kahler\ superspace.  The relation
between them will be discussed in detail later.  If we further
assume that there is an internal gauge group, then we must
introduce yet another one-form gauge connection $\A_a\,^b = dz^M
\A_{Ma}\,^b$.  Solving the Bianchi identities now introduces a
new superfield strength, $W_\a\,^a$.

Using these superfields, one can write down the tree-level
superfield \lag\ in this superspace.  Again, it consists of three
parts, each specified by a fundamental function.  The first part
is the supergravity-matter kinetic energy term given by
\beq \L_0 = -3 \kappa^{-2} \int d^4 \theta E e^{-\frac{{\kappa}^2}{3}
K(\Phi_i, \Phi_i^{\dagger}) } \eeq 
The second part is the potential energy term given by
\beq \L_{PE} = \frac{1}{2} \int d^4 \theta \frac{E}{R} W(\Phi_i)
+ h.c. \eeq 
Finally, the Yang-Mills \lag\ is
\beq \L_{YM} = \frac{1}{8} \int d^4 \theta \frac{E}{R}
f(\Phi_i)_{ab} \; W^{\a a} W_\a\,^b + h.c. \eeq 
Note that $E$ is the superdeterminant in conventional $SL(2,C)$
superspace and is not identical to the superdeterminant in
\Kahler\ superspace discussed earlier.  Now, the total tree-level
\lag\ possesses, by construction, both gauge and Lorentz
invariance, as did the \Kahler\ superspace \lag.  However, it is
clear that under the \Kahler\ transformation
${\kappa}^2 K \ra {\kappa}^2 K + F +
\bar{F}$ (4.2), for example, is not invariant.  That is, the
conventional supergravity-matter \lag\ does not possess pure
\Kahler\ symmetry.  Is there anything that replaces it?  After a
short discussion, we will show that indeed
there is, as is well known
\cite{1,2,90}.

The conventional torsion constraints are left invariant
\cite{18} under a set of super-scaling transformations
called super-Weyl or Howe-Tucker transformations.  They change
the forms of superspace as follows
\beqa E_M \,^a &\ra& e^{\Sigma + \bar{\Sigma}} E_M\,^a \nonumber
\\ E_M\,^\a & \ra & e^{2 \bar{\Sigma} - \Sigma} \left( E_M\,^\a +
\frac{i}{2} E_M\,^b \left( \epsilon \sigma_b \right)^\a
\,_{\dota} \bar{\D}^{\dota} \bar{\Sigma} \right) \nonumber \\
E_{M \dota} &\ra& e^{2 \Sigma - \bar{\Sigma}} \left( E_{M \dota}
+ \frac{i}{2} E_M\,^b \left( \epsilon \bar{\sigma}_b
\right)_{\dota} \,^\a \;\D_\a \Sigma \right) \eeqa 
where $\Sigma$ and $\bar{\Sigma}$ are superfield parameters
subject to the chirality conditions
\beqa \bar{\D}^{\dota} \Sigma &=& 0 \nonumber \\
\D_\a \bar{\Sigma} &=& 0 \eeqa 
It is not hard to show that under (4.5)
\beq E \ra E \; e^{2 (\Sigma + \bar{\Sigma})} \eeq 
Chiral superfields and $K(\Phi_i, \Phi_i^{\dagger})$ are invariant
under super-Weyl transformations.  It is clear
that (4.2), for
example, and, hence, the conventional supergravity-matter \lag\
is not invariant
under pure super-Weyl rescalings.  However, under combined
\Kahler\ and super-Weyl transformations (4.2) transforms as
\beqa \L'_0 &=& - 3 \kappa^{-2} \int d^4 \theta E' e^{-
\frac{{\kappa}^2}{3} K'} \nonumber \\
&=& - 3 \kappa^{-2} \int d^4 \theta E e^{2 (\Sigma +
\bar{\Sigma})} e^{-\frac{{\kappa}^2}{3} K'}
e^{-\frac{1}{3} (F + \bar{F})}
\eeqa 
It is clear that if we take
\beqa \Sigma &=& \frac{1}{6} F \nonumber \\ \bar{\Sigma} &=&
\frac{1}{6} \bar{F} \eeqa 
then \lag\ (4.2) will be invariant.  Similarly, if one demands
that under \Kahler\ transformations, the superpotential transform
as
\beq W \ra e^{-F} W \eeq 
then, using the fact that under super-Weyl rescalings
\beqa \delta R & = & -2 (2 \Sigma - \bar{\Sigma}) R - \frac{1}{4}
\bar{\D}^2 \bar{\Sigma} \nonumber \\
W_\a & \ra & e^{-3 \Sigma} W_\a \eeqa 
it follows that $\L_{PE}$, (4.3), and $\L_{YM}$, (4.4), will
also be invariant under combined super-Weyl-\Kahler\
transformations with $\Sigma$ given in (4.9).  We conclude,
therefore that instead of possessing pure \Kahler\ symmetry, the
conventional supergravity-matter tree-level \lag\ exhibits a
mixed super-Weyl-\Kahler\ invariance \cite{1,2,90}.

Before considering the component field \lag, it is very helpful
to first reexpress $\L_0$ in (4.2) and $\L_{PE}$ in (4.3)
in a way that exposes, at the
superfield level, the vertices relevant to the anomaly
calculation.  As in K\"{a}hler superspace, $\L_{YM}$ in
(4.4) does not contribute to K\"{a}hler anomalous one-loop
irreducible graphs and will, henceforth, be ignored.
To exhibit the relevant vertices,
we perform a background field-
quantum field splitting of the chiral matter superfields
as well as of the supergravity superfields in such a way
that, all relevant
quantum superfields are invariant under
super-Weyl-K\"{a}hler transformations of the background superfields.
This proceeds as follows.
First perform a background field-quantum
field splitting of the chiral matter superfields as
\beq \Phi_i = \Phi_{ci} + \phi_i \eeq 
where $\Phi_{ci}$ is a background superfield and $\phi_i$ is the
fluctuating quantum superfield.  Henceforth, for simplicity of
notation, we will suppress the index $i$.
Recalling that $\Phi$ does not transform under
super-Weyl-K\"{a}hler transformations, and demanding that
$\Phi_c$ also not transform, it follows that under
super-Weyl-K\"{a}hler transformations
\beq  \delta \phi = 0 \eeq
To proceed further, we must discuss the background field-quantum
field splitting of the supergravity superfields.  It is well
known \cite{19}
that the superdeterminant $E$ can be written in terms of
a set of supergravity prepotentials $H^A$ and $\rho$, where
$\rho$ is a chiral superfield.  In terms of these prepotentials
$E$ factors into
\beq E = \hat{E} [H^A] e^{\rho + {\bar {\rho}}} \eeq
where $\hat{E}$ is a complicated function of $H^A$ whose exact
form is irrelevant for this paper.  What is relevant is that
under super-Weyl transformations $\hat{E}$ is invariant.
Superprepotential $\rho$, on the other hand, transform as
\beq  \rho \ra \rho + 2 \Sigma   \eeq
under super-Weyl transformations, which correctly
reproduces (4.7).  Recall that under super-Weyl-K\"{a}hler
transformations $\Sigma$ satisfies (4.9) and, therefore,
that
\beq   \rho \ra \rho + \frac{1}{3} F \eeq
We now perform a background field-quantum field splitting
of $\rho$ as
\beq  \rho = \rho_c + \rho_q    \eeq
where $\rho_c$ is a chiral background superfield and
$\rho_q$ a fluctuating chiral quantum superfield.  Using
(4.16), and demanding that
\beq \rho_c \ra \rho_c + \frac{1}{3} F_c  \eeq
it follows from (4.17) that under super-Weyl-K\"{a}hler
transformations
\beq  \rho_q  \ra  \rho_q + \frac{1}{3} F_{\phi c} \phi
+ \frac{1}{6} F_{\phi \phi c} {\phi}^2 + \cdots  \eeq
where the subscript $\phi$ means differentiation with
respect to these fields and subscript $c$ means
evaluation at the classical background
superfield.
Note that this transformation mixes quantum gravity superfields
with quantum matter superfields.  Amongst other things, this
implies that the relevant part of the
naive path integral measure, $[ \D \phi]
[\D {\phi}^{\dagger}] [\D \rho_q] [\D {\rho}^{\dagger}_q]$,
is not invariant under super-Weyl-K\"{a}hler transformations.
This can be cured by introducing the appropriate Jacobian
determinant but there is an easier approach, which we will now
follow.  This consists of defining a new chiral quantum
supergravity superfield $\rho'_q$ by
\beq \rho'_q = \rho_q + \frac{1}{3} (ln W)_{\phi c} \phi
+ \frac{1}{6} (ln W)_{\phi \phi c} {\phi}^2 + \cdots \eeq
Combining (4.10), (4.19) and (4.20), it follows that under
super-Weyl-K\"{a}hler transformations
\beq \delta \rho'_q = 0  \eeq
In this new variable, the relevant part of the
path integral measure,
$[\D \phi] [\D {\phi}^{\dagger}] [\D \rho'_q]
[\D {\rho'_q}^{\dagger}]$ with unit Jacobian, is indeed
invariant under super-Weyl-K\"{a}hler transformations.
Henceforth, we
will use $\phi$ and $\rho'_q$ as the fluctuating quantum
superfields.

We will first display the couplings
of uncharged quantum matter
superfields $\phi$ to background
supergravity and background matter fields.
Expanding $K(\Phi, {\Phi}^{\dagger})$ to
quadratic order around classical background $\Phi_c$ gives
\beqa K &=& K_c + K_{\phi c} \phi + K_{{\phi}^{\dagger} c}
{\phi}^{\dagger} \nonumber \\
&& + K_{\phi {\phi}^{\dagger} c} {\phi}^{\dagger} \phi
\nonumber \\
&& + \frac{1}{2} K_{\phi \phi c} \phi^2 +
\frac{1}{2}
K_{{\phi}^{\dagger} {\phi}^{\dagger} c}
{{\phi}^{\dagger}}^2 + \ldots  \eeqa 
where, again,
the subscript $\phi$ or ${\phi}^{\dagger}$ means differentiation
with respect to these fields and subscript $c$ means
evaluation at the classical background superfields.
A similar expansion holds for the
holomorphic superpotential $W$.
Substitute these expansions into Lagrangians
(4.2) and (4.3) and use (4.14), (4.17) and (4.20) to express
the superdeterminant.  Setting the terms linear in $\phi$
to zero leads to the following
equation of motion for background field $\Phi_c$
\beq \left( \bar{\D}^2 - 8 R \right)_c
\left[ e^{-\frac{{\kappa}^2}{3} K_c}
G_{\phi c} \right] = 0 \eeq 
where $G$ denotes the K\"{a}hler invariant combination
$G = K + ln W + ln \bar{W} $.  Note that the equation of motion
(4.23) is manifestly covariant under super-Weyl-K\"{a}hler
transformations of the background superfields.  Also note that
we have ignored a possible contribution of Lagrangian (4.4)
to the equation of motion for $\Phi_c$, (4.23).  As stated
earlier, such matter couplings do not lead to super-Weyl-K\"{a}hler
anomalous contributions steming from irreducible graphs.
The resulting Lagrangian now becomes
\beqa \L_0 + \L_{PE} &=& - 3 \kappa^{-2} \int d^4 \theta
E_c e^{-\frac{\kappa^2}{3}  K_c}  +
( \frac{1}{2} \int d^4 \theta \frac{E_c}{R_c} W_c + h.c. )
\nonumber \\
&& + \int d^4 \theta E_c e^{-\frac{\kappa^2}{3} K_c}\left[ \left(
g_{\phi {\phi}^{\dagger}} - \frac{\kappa^2}{3} G_{\phi}
G_{{\phi}^{\dagger}} \right)_c \right.
{\phi}^{\dagger} \phi \nonumber \\
&& + \frac{1}{2} \left( G_{\phi \phi } -
\frac{\kappa^2}{3} G_{\phi} G_{\phi}
\right)_c \phi^2 \nonumber \\
&& \left. + \frac{1}{2} \left( G_{{\phi}^{\dagger} {\phi}^{\dagger}}
- \frac{\kappa^2}{3} G_{{\phi}^{\dagger}}
G_{{\phi}^{\dagger}} \right)_c {{\phi}^{\dagger}}^2 \right]
+ \ldots \eeqa 
where we have dropped all terms depending on ${\rho}'_q$,
since we will not compute gravitational radiative corrections
in this paper.  For the same reason we have simply replaced
$H^A$ by $H^A_c$ in $\hat{E}$.
Lagrangian (4.24) displays the
couplings of uncharged quantum
superfields $\phi$ to supergravity and background matter fields.
We now proceed to display the couplings of charged
quantum superfields $\phi$.
We will throughout the paper
assume that
all chiral matter superfields that carry non-vanishing charge
under some Yang-Mills gauge group have a vanishing
background.
Then, the corresponding \lag\ reads
\beqa \L_0 &=& - 3\kappa^{-2} \int d^4 \theta
E_c e^{-\frac{\kappa^2}{3} K_c}  \nonumber \\
&& + \int d^4 \theta E_c e^{-\frac{\kappa^2}{3} K_c} g_{\phi
{\phi}^{\dagger} c} {\phi}^{\dagger} e^V \phi + \ldots
\eeqa
Note that terms proportional to ${\phi}^2$ and to
${{\phi}^{\dagger}}^2$ do not appear.  Both Lagrangians (4.24)
and (4.25) are manifestly invariant under super-Weyl-K\"{a}hler
transformations (4.9) of the background superfields.
Also note that the
terms in (4.24) proportional to $\phi^2$ and ${{\phi}^{\dagger}}^2$ come
multiplied by the super-Weyl-K\"{a}hler invariant, but
background field dependent metrics $G_{\phi \phi c}
- \frac{\kappa^2}{3}\;\; G_{\phi c} G_{\phi c}$ and $G_{{\phi}^{\dagger}
{\phi}^{\dagger} c} -\frac{\kappa^2}{3}\;\; G_{{\phi}^{\dagger} c}
G_{{\phi}^{\dagger} c}$, respectively.

We are now ready to expand Lagrangians
(4.24) and (4.25) into
component fields using the standard techniques.
The relevant
part of the \lag\ given in (4.24) reads, to lowest order
in the background fields,
\beqa \L_0/e &=& \ldots + \left( g_{ A A^{\dagger}}
- \frac{\kappa^2}{3} G_A
G_{A^{\dagger}}  \right)_c e^{- \frac{\kappa^2}{3}
K_c \left. \right|}\nonumber \\
& & \left[ \left\{ \frac{1}{6} \bar{A} A \left( \R -
\frac{{\kappa}^2}{2}(4 K \left.
\right|_{\theta^2 \bar{\theta}^2} + \Box_0 K \left.
\right| ) \right)_c \right. \right. \nonumber \\
& & \;\;\; - g^{mn} \tilde{\D}_m A \tilde{\D}_n \bar{A} \nonumber
\\
& & \left.- \frac{i}{2}  \left( \chi \sigma^m \tilde{\D}_m
\bar{\chi} - \left( \tilde{\D}_m \chi \right) \sigma^m \bar{\chi}
\right) \right\} \nonumber \\
& & + \left\{ \frac{ \left( G_{ A A}  -
\frac{\kappa^2}{3} G_A G_A
\right)_c}{ \left( g_{ A A^{\dagger}} -
\frac{\kappa^2}{3} G_A G_{A^{\dagger}}
\right)_c} \left[ \frac{1}{12} \left( \R -
\frac{{\kappa}^2}{2} (4 K
\left|_{\theta^2 \bar{\theta}^2} \right.\right. \right. \right.
\nonumber \\
& & \left. \left. \left. \left. + \Box_0 K \left| )
\right. \right) + \frac{1}{6} i \partial^m b_m \right]_c A^2 +
h.c. \right\} \right] \nonumber \\
& & + \ldots \eeqa 
where the covariant derivatives are
\beq \tilde{\D}_m A^i = ( \partial_m - \frac{i}{3} \left( b_m -
i2 \kappa^2 a_m \right) ) A^i \eeq 
and
\beq
\tilde{\D}_m \chi^i = ( \partial_m  - \omega_m - \frac{2}{3} {\kappa}^2
c_m ) \chi^i
\eeq 
and where
\beq c_m = a_m - {\kappa}^{-2} \frac{i}{4} b_m \eeq 
Note that $b_m, a_m, \omega_m$ and $c_m$ are all
evaluated at classical background field values.  We omit
the subscript $c$ to prevent a confusing proliferation
of notation.
The appearance of the connections $b_m$ and $a_m$ in (4.26) and,
in particular, the appearance of the exact combination
$c_m$ requires
some explanation.  In the conventional superspace $\chi^i$ does not
transform under pure \Kahler\ transformations.  However, under
super-Weyl transformations $\chi^i$ transforms as
\beq \delta \chi^i = - \left( 2 \bar{\Sigma} - \Sigma \right)
\left| \chi^i \right. \eeq 
Similarly, under super-Weyl transformations it can be shown that
\beq  \delta b_m \left. = - 3 i \partial_m \left( \bar{\Sigma} -
\Sigma \right) \right| \eeq 
Note that this variation of $b_m$ does not compensate the
variation of $\chi^i$.  It need not, since the \lag\ is not
purely super-Weyl invariant.  If we now also perform a \Kahler\
transformation, $a_m$ transforms as in (2.20).  However, if we
further demand that (4.9) hold, we then find from (2.20) and (4.31)
that
\beq \left. \delta c_m = {\kappa}^{-2} \frac{3}{8}
\partial_m \left( F - \bar{F} \right) \right| \eeq 
which exactly compensates (4.30), as it must.  That is, the exact
combination $c_m = a_m - {\kappa}^{-2} \frac{i}{4} b_m$
appearing in the
covariant derivative (4.28) acts as a connection which insures
invariance under mixed super-Weyl-\Kahler\ transformations.  Also
note that the combination $b_m - 2i \kappa^2 a_m$ appearing in
the covariant derivative $\tilde{\D}_m A$ in (4.27) is invariant
under mixed  super-Weyl-\Kahler\ transformations, as it must,
since the component scalar field $A$ does not transform under
mixed super-Weyl-\Kahler\ transformations.

The component expansion of \lag\ (4.25) reads, to lowest order
in the background fields,
\beqa \L_0/e & = & \ldots g_{A A^{\dagger} c}
e^{- \frac{\kappa^2}{3} K_c \left| \right. } \left\{ \frac{1}{6}
\bar{A} A \left( \R - \frac{{\kappa}^2}{2}
(4 K \left|_{\theta^2 \bar{\theta}^2}
\right.
+ \Box_0 K \left| ) \right. \right)_c \right. \nonumber \\
&& - g^{mm} \tilde{\D}_m A \D_n \bar{A} + \bar{A} T^{(r)} A
\D^{(r)} \nonumber \\
&& - \frac{i}{2} \left. \left( \chi \sigma^m \tilde{\D}_m
\bar{\chi} - \left( \tilde{\D}_m \chi \right) \sigma^m \bar{\chi}
\right) \right\}\nonumber \\ && + \ldots \eeqa 
where the covariant derivatives are
\beqa  \tilde{\D}_m A^i &=& ( \partial_m + i v_m^{(r)}
T^{(r)}
-\frac{i}{3} \left( b_m - 2 i \kappa^2 a_m \right) ) A^i
\nonumber \\
\tilde{\D}_m \chi^i &=&  ( \partial_m - \omega_m + i
v_m^{(r)} T^{(r)} - \frac{2}{3} {\kappa}^2 c_m ) \chi^i \eeqa 
Again, $v_m^{(r)}, b_m, a_m, \omega_m$ and $c_m$ are to be
evaluated at classical background field values.
Note that many terms appearing in (4.26) do not appear in (4.33)
since we have set charged background fields to zero.

\section{Super-Weyl-\Kahler\ Anomalies in Conventional Superspace
- The Yang-Mills Case}

\setcounter{equation}{0}

\hspace*{.3in}
We are now poised to compute mixed super-Weyl-\Kahler-gauge and
Lorentz anomalies, in exact analogy with the mixed \Kahler-gauge
and Lorentz anomalies discussed in section 3.  A key difference
in these calculations is that now the anomalous fermionic
current couples
to connection $c_m$ rather than to $a_m$.  We will, in this
section, consider the mixed super-Weyl-\Kahler-gauge anomaly
only.  The mixed gravitational anomalies will be discussed in the
next section.  We will work to lowest order in the
supergravity-K\"{a}hler background fields.
As before, we consider a set of matter chiral
superfields, $\Phi_i$, the subset of which with non-vanishing
gauge charges so chosen that they are free of gauge and Lorentz
anomalies.  The relevant part of the supergravity-matter \lag\ is
given in (4.33), with the covariant derivatives given in (4.34).
The graph in Figure~2a gives the fermionic contribution to the
mixed super-Weyl-\Kahler-gauge anomaly.  This graph can be
evaluated \cite{17}
and the associated effective \lag\ is found to be
\beq \L_\chi = \frac{2i}{3} \frac{\kappa^2}{(4\pi)^2} \; Tr \;
F_{mn} \tilde{F}^{mn} \frac{1}{{\Box}_0} \partial^p  c_p.\eeq 
In this expression, and in all results for the remainder of
this paper, all fields are to be evaluated at their classical
background values.  We omit the subscript $c$ to avoid a
confusing proliferation of notation.
Implied in (5.1), as everywhere else in this section, is a sum over
all charged matter multiplets ${\phi}_i$.
It is readily seen that (5.1) is anomalous under
super-Weyl-\Kahler\ transformations.  Using (4.32), we find that
\beq \delta_{SW-K} \L_\chi = - \frac{1}{2} \; \frac{1}{(4\pi)^2}
\; Im \; F \; Tr \; F_{mn} \tilde{F}^{mn} \eeq 
which, since it does not vanish, implies that the super-Weyl-\Kahler\
symmetry is broken.  Now $\L_\chi$ by itself is not
supersymmetric.  Before trying to find the one-loop graphs that
will supersymmetrize (5.1), it is worth noting that there is a
unique superfield expression whose highest component contains
$\L_\chi$.  This expression is given by
\beq \L = \frac{1}{4(4\pi)^2} \int d^2 \theta \; Tr \; W^\a W_\a
\frac{1}{\Box_0} \left( 4 R^{\dagger} - \frac{\kappa^2}{3} D^2 K
\right) + h.c. \eeq 
Expanding (5.3) out into component fields yields
\beqa \L &=& \frac{2i}{3} \; \frac{{\kappa}^2}{(4 \pi)^2}
Tr \; F_{mn} \tilde{F}^{mn} \frac{1}{\Box_0} \partial^p c_p
\nonumber \\
&& \;\;\;\;\; - \frac{1}{12} \frac{{\kappa}^2}{(4 \pi)^2}
Tr \; F_{mn} F^{mn}
\frac{1}{\Box_0} \left( 4 K |_{\theta^2 \bar{\theta}^2} +
\Box K | \right)  \nonumber \\
&& \;\;\;\;\; - \frac{1}{12(4 \pi)^2}
Tr \; F_{mn} F^{mn}
\frac{1}{\Box_0} \R  + \ldots \eeqa 
where $\ldots$ refers to terms containing fermionic and
$K|_{\theta^2 \bar{\theta}^2}$ and $K |$ are given in (3.5).
Can one find graphs that will produce the second and third terms
on the right hand side of (5.4)?  Unlike the case in \Kahler\
superspace, the answer now is affirmative.  To begin with, we
consider the graph shown in Figure~2b where the scalar fields
$A^i$ run around the loop.  The top vertex originates through the
quadratic coupling of $A^i$ to $4K\left|_{\theta^2
\bar{\theta}^2}\right. + \Box_0 K \left| \right.$
in the component field \lag\ (4.33).  This is precisely the
coupling that was conspicuously absent in \Kahler\ superspace.
Evaluation of this graph leads to the following additional
contribution to the effective action
\beq \L_A = - \frac{{\kappa}^2}{12(4\pi)^2} \; Tr \; F_{mn} \; F^{mn}
\frac{1}{\Box_0} \left( 4 K\left|_{\theta^2 \bar{\theta}^2}
\right. + \Box K \left| \right. \right) \eeq 
precisely the second term on the right-hand side of (5.4).  The
origin and meaning of the third term in (5.4) is a bit more
subtle.  The relevant graphs are shown in Figure~2c.  The top
vertices originate through the fermion and scalar kinetic energy
terms in (4.33).  Evaluation of these graphs leads to the
following contribution to the effective action
\beq \L_{\chi,A} = - \frac{1}{3(4\pi)^2} \; Tr \; F_{mn} F^{mn}
\frac{1}{\Box_0} \left( \frac{1}{4} \R \right).\eeq 
This is exactly the third term on the right-hand side of (5.4).
There are two issues that must be addressed as regards this term.
First of all, one might wonder why we did not consider such
graphs in our discussion of \Kahler\ superspace.  The reason is
that in \Kahler\ superspace, $\R$ does not transform under
\Kahler\ transformations and, hence, these graphs are irrelevant
for the discussion of the \kahler\ anomaly.  In the conventional
superspace of this section, however, $\R$ does transform under
the combined super-Weyl-\kahler\ transformations and, therefore,
these graphs are relevant.  Secondly, we want to point out that
the graphs in Figure~2c are closely related to the gauge
two-point function and the associated trace anomaly
\cite{29,30,21}.  The gauge two-point function is computed from
the graphs in Figure~3.  Prior to renormalization these graphs
are evaluated to be
\beq \L_2 = - \frac{1}{(4 \pi)^2} \; \frac{1}{4} \left(
\frac{1}{\epsilon} - \ln \frac{\Box_0}{2\pi {\mu}^2} \right) Tr
\; F_{mn} F^{mn} \eeq 
using dimensional regularization.  The energy momentum tensor is
defined by
\beq T_{mn} = \frac{2}{\sqrt{-g}}\;\; \frac{\delta S}{\delta
g^{mn}}.\eeq 
Varying the action associated with (5.7), one finds that the
trace of $T_{mn}$ is finite in four-dimensions and given by
\beq T^m\,_m = - \frac{1}{2(4 \pi)^2} Tr \; F_{mn} F^{mn}.\eeq 
This result occurs because the variation of $\sqrt{-g} \;Tr \;
F_{mn} \; F^{mn}$ in $4-2\epsilon$ dimensions is non-vanishing and
proportional to $\epsilon$.  This multiplies the
$\frac{1}{\epsilon}$ pole and yields the finite result in (5.9).
Now, this is true for the non-renormalized theory.  However, when
(5.7) is renormalized, the pole $1/\epsilon$ is removed, and it
is clear that the renormalized $T^m\,_m=0$ in four-dimensions when
evaluated from the graph in Figure 3.
Now add graphs~2c to the renormalized gauge two-point effective
\lag.  The result in four-dimensions is
\beq \L_2 + \L_{\chi ,A} = \frac{1}{4(4\pi)^2} \ln \left[
\frac{\Box_0}{2\pi {\mu}^2} \left(1 - \frac{1}{3\Box_0} \R
\right) \right] Tr \; F_{mn} F^{mn}.\eeq 
It follows from (5.8) that the associated trace of the energy
momentum tensor is
\beq T^m\,_m = \frac{1}{2(4 \pi)^2} Tr \; F_{mn} F^{mn} \eeq 
which is, up to the sign, the same as in (5.9).  The right-hand
side of (5.11) is the correct expression for the one-loop trace
anomaly \cite{21} of gauge indexed chiral multiplets,
each of which has two real scalars and one Weyl fermion.
Note that it arises entirely from
varying $g_{mn}$ in the three-point effective \lag\ (5.6) and not
in the renormalized two-point \lag.  Hence, $\L_{\chi A}$ is the
term in the renormalized effective \lag\ that gives rise to the
one-loop trace anomaly.

We conclude, then, that the one-loop
three-point graphs shown in Figure~2a, b, and c give rise to the
supersymmetric effective \lag\ (5.4) and, equivalently, to the
superfield expression for the \lag\ (5.3).  Under
super-Weyl-\kahler\ transformations
\beqa {\kappa}^2 \delta K &=& F + \bar{F} \nonumber \\
\delta R^{\dagger} &=& - \frac{1}{24} D^2 F \eeqa 
it follows that
\beq \delta \L = \frac{1}{2(4\pi)^2} \int d^2 \theta \; Tr \;
W^\a W_\a F + h.c. \eeq 
which is non-vanishing.  Hence, expression (5.3) represents the
supersymmetric mixed super-Weyl-\kahler-gauge anomalous term in
the effective action.  We point out that expression (5.3) could
also have been obtained by direct three-point supergraph
calculations using the gauge and supergravitational prepotential
formalism \cite{19,8}.  We close this section by showing how
to transform the anomalous \lag\ (5.3), obtained in the
conventional superspace formulation, over into a result in the
\kahler\ superspace formulation.  Both formulations of
supergravity-matter are related by particular
superfield rescalings of the
underlying torsion constraints \cite{6}.  This implies,
among other things, that the superfields $R$ and $G_{\a \dota}$
of the conventional superspace formulation are related to the
superfields $R$ and $G_{\a\dota}$ of the \kahler\ superspace
formulation by \cite{6}
\beqa R \ra R & - & \frac{\kappa^2}{24} {\bar{D}}^2 K \nonumber \\
G_{\a\dota} \ra G_{\a \dota} &-& \frac{\kappa^2}{12} \left[
D_\a, \bar{D}_{\dota} \right] K \eeqa 
where we work to linearised level in $K$ only.  It follows
from (5.14) that
\beq b_m \ra b_m + 2 i \kappa^2 a_m \eeq 
Applying (5.14) to the anomalous \lag\ (5.3) yields
\beqa \L &=& \frac{1}{(4 \pi)^2} \int d^4\theta \; Tr \;W^\a W_\a
\frac{1}{\Box_0} R^{\dagger} \nonumber \\
&& - \frac{1}{8} \;\; \frac{{\kappa}^2}{(4 \pi)^2}
\int d^4 \theta \; Tr \;
W^\a W_\a \frac{1}{\Box_0} D^2 K + h.c. \eeqa 
Note that \lag\ (5.16) is given in terms of \kahler\ superspace
superfields $W^{\alpha}$, $R^{\dagger}$, $K_c$.
In K\"{a}hler superspace the superfield $R^{\dagger}$ transforms,
according to (2.6),
homogeneously with K\"{a}hler charge $\omega_K = -2$ under
K\"{a}hler transformations.  It follows that the
term proportional to $R^{\dagger}$ in (5.16) is invariant
under K\"{a}hler transformations.  Hence, it can be discarded
and the K\"{a}hler anomaly is entirely given by the second term
in (5.16).  That is
\beq \L^{anom} = - \frac{1}{8} \;\; \frac{{\kappa}^2}{(4 \pi)^2}
\int d^4 \theta \; Tr \;
W^\a W_\a \frac{1}{\Box_0} D^2 K + h.c. \eeq 
Comparing this expression
against \lag\ (3.3) shows that both agree!  Hence, it
follows that a consistent way of generating supersymmetric
one-loop
results in the \kahler\ superspace formulation is to first
perform manifestly supersymmetric calculations in the
conventional formulation, and then to transform them over into
the \kahler\ superspace formulation by using the superfield
rescalings of the torsion constraints given in \cite{6}.
We will, in the next section, apply this strategy to the
computation of the mixed supersymmetric \kahler-Lorentz anomaly
in \kahler\ superspace.

Finally, recall that $Tr T^{(r)} T^{(s)} = \sum_{H} g^2_H
\sum_{R} c^H_R \delta^{(r)(s)}$, where each $H$ is a factor
group of the total gauge group, $g_H$ is the gauge coupling
parameter associated with $H$ and $R$ sums over various
representations of multiplets within each factor gauge group.
It follows that (5.17) can be written as
\beq \L = - \frac{1}{8} \;\; \frac{{\kappa}^2}{(4 \pi)^2}
\sum_H g_H^2 \int d^4 \theta
(W_{\alpha}^{H (r)})^2 \frac{1}{\Box_0} D^2
[\sum_R c_R^{H} K ] + h.c. \eeq 
In this paper, we have, for simplicity, not discussed K\"{a}hler
anomalies with internal vector supermultiplet loops.
Similarly, we have ignored anomalies involving the sigma-model
Christoffel connection \cite{22,10,7,9}.  It seems clear that these
calculations will proceed in a manner similar to that
discussed above.  Assuming this to be the case, the result
(5.18) can be easily extended to give the complete supersymmetric
K\"{a}hler and sigma-model anomalies for both chiral
and vector superfield internal loops.  The final
result is \cite{10,7,9}
\beq \L = \frac{1}{8} \;\; \frac{{\kappa}^2}{(4 \pi)^2}
\sum_H g_H^2 \int d^4 \theta
(W_{\alpha}^{H (r)})^2 \frac{1}{\Box_0} D^2
[c_V^H K  + \sum_R c_R^{H} ( 2 ln det g_R^H - K) ] + h.c. \eeq 

\section{Super-Weyl-\kahler\ Anomalies in Conventional Superspace
- The Gravitational Case}

\setcounter{equation}{0}

\hspace*{.3in} We now turn to the computation of the mixed
super-Weyl-\Kahler-Lorentz anomaly.  We will, again, work to lowest
order in supergravity-K\"{a}hler background fields.  All results
presented in this section are for one chiral matter supermultiplet
running in the loop.  The generalization of these results to $N$
chiral quantum matter supermultiplets is obtained by multiplying
all the results of this section by $N$.
The relevant part of the
supergravity-matter \lag\ is given in (4.26) with the covariant
derivatives given in (4.27) and (4.28).  The graph in Figure~4a
gives the fermionic contribution to  the mixed
super-Weyl-\kahler-Lorentz anomaly.  This graph can be evaluated
\cite{17}
and the associated effective \lag\ is found to be
\beq \L'_\chi = - \frac{2i{\kappa}^2}{3} \;\; \frac{1}{24} \;\;
\frac{1}{(4\pi)^2} \R_{mn a}\,^b \tilde{\R}^{mna}\,_b
\frac{1}{\Box_0} \partial^p c_p.\eeq 
Proceeding by analogy with the previous case of the mixed gauge
anomaly, we now also evaluate three-point graphs with scalar
fields $A$, $\bar{A}$ running in the loop.  Unlike the previous
case, however, there are now $A^2$ and $\bar{A}^2$ as well as
$\bar{A}A$ vertices to consider.  These vertices fall into two
categories, the ``conformal'' vertices $A\bar{A}$ and the
``conformal breaking'' vertices $A^2$ and $\bar{A}^2$.  The
conformal vertices arise from the ${\phi}^{\dagger} \phi$ term in \lag\
(4.24), which is invariant under the conformal transformations.
These are defined by arbitrary super-Weyl transformations, as
given in (4.7), accompanied by the superfield rescalings of the
quantum fields $\phi$ and ${\phi}^{\dagger}$, given by
\beqa \phi &\ra& e^{-2 \Sigma} \phi \nonumber \\
{\phi}^{\dagger} &\ra& e^{-2 \bar{\Sigma}} {\phi}^{\dagger}. \eeqa 
{}From (6.2) it follows immediately that the $\phi^2$ and
${{\phi}^{\dagger}}^2$ terms in \lag\ (4.24) are not invariant under the
conformal transformations.  Hence, the $A^2$ and $\bar{A}^2$
vertices in the component \lag\ (4.26) are of the conformal
breaking type.

We begin by evaluating three-point graphs constructed from the
conformal vertices only.  These graphs are shown in Figures~4b
and 4c.  The graph in Figure~4b has scalars running around the
loop.  We find that its contribution to the effective \lag\ is
\beqa \L'_A &=& \frac{\kappa^2}{12 \cdot 24\cdot (4\pi)^2} \left(
d_M C_{mnpq} C^{mnpq} - (d_M - 1) GB \right) \frac{1}{\Box_0}
\nonumber \\
&& \;\;\;\;\; \x \left( 4K \left|_{\theta^2 \bar{\theta}^2}
\right. + \Box_0 K \left| \right. \right) \eeqa 
where $GB$ denotes the Gauss-Bonnet combination
\beq GB = C_{mnpq} C^{mnpq} - 2 \R_{mn} \R^{mn} + \frac{2}{3}
\R^2. \eeq 
and we have not computed coefficient $d_M$ as it is irrelevant
to the physics we wish to discuss.
Note that the only Lorentz invariant gravitational combinations
appearing in (6.3) are the square of the Weyl tensor, $C_{mnpq}$,
and the  Gauss-Bonnet combination $GB$.  Also
note that these are the only
quadratic
gravitational combinations that transform homogeneously under
arbitrary Weyl-rescalings of the space-time metric $g_{mn}$.
The fact that only these two combinations appear in (6.3) is a
consequence of the conformal properties of the $\bar{A}A$
vertices of the graph shown in Figure~4b.  Similarly, the fermion
and scalar loop graphs of Figure~4c can be computed, and their
contribution is found to be \cite{29,30}
\beq \L'_{\chi, A} = \frac{1}{6 \cdot 24 \cdot (4\pi)^2} \left(
C_{mnpq} C^{mnpq} - \frac{1}{2} GB \right) \frac{1}{\Box_0} \R.
\eeq 
Once again, only the two gravitational combinations compatible
with the conformal symmetry of the vertices appear in (6.5).

As before in the gauge case, the contribution of (6.5) to the
renormalized trace of the energy momentum tensor can be evaluated
using (5.8).  We find that
\beq T^{conf} \,^{m} \,_m = - \frac{1}{24} \; \frac{1}{(4\pi)^2} \left(
C_{mnpq} C^{mnpq} - \frac{1}{2} GB  \right) \eeq 
which is the well known gravitational contribution to the
one-loop trace anomaly from conformal scalars and Weyl-fermions
\cite{29,33}.  Note that (6.6) can be put in the form
\beq T^{conf} \,^{m} \,_m = - \frac{1}{24} \; \frac{1}{(4\pi)^2} \left(
3b C_{mnpq} C^{mnpq} - b' GB  \right) \eeq 
where
\beqa b &=& \frac{1}{15} (N_S + 3 N_F)  \nonumber\\
b' &=& \frac{1}{15} (N_S + \frac{11}{2} N_F) \eeqa
and $N_S(=2)$ is the number of scalars fields and $N_F(=1)$
is the number of Weyl fermions.
Adding (6.1), (6.3) and (6.5) we find that
\beqa \L^{conf}
&=& - \frac{2i{\kappa}^2}{3 \cdot 24 \cdot (4 \pi)^2}
\R_{mna}\,^b \; \tilde{\R}^{mn\;\;a} \!\!\!_b \frac{1}{\Box_0}
\partial^p c_p \nonumber \\
&& + \frac{1}{12 \cdot 24 \cdot (4\pi)^2} C_{mnpq} C^{mnpq}
\frac{1}{\Box_0} (6b \R + d_M \kappa^2 \left[ 4 K|_{\theta^2
\bar{\theta}^2} + \Box_0 K| \right] )         \nonumber \\
&&- \frac{1}{12 \cdot 24 \cdot (4\pi)^2}
GB \frac{1}{\Box_0}
\left( 2b' \R + (d_M - 1)\kappa^2 \left[ 4 K \left|_{\theta^2
\bar{\theta}^2} \right. + \Box_0 K \left| \right. \right]
\right) \eeqa 
Using the identities \cite{25}
\beqa \frac{1}{16} (C_{mnpq} C^{mnpq} - i
\R_{mna}\,^b \; \tilde{\R}^{mn\;\;a} \!\!\!_b ) =
W^2_{\alpha \beta \gamma} |_{{\theta}^2} \nonumber \\
GB = \left(
8 W^2_{\a\b\gamma} + \left( \bar{\D}^2 - 8 R\right) \left(
G^2_{\a\dota} - 4 R^{\dagger} R \right) \right)|_{{\theta}^2} + h.c.
\eeqa 
it follows that (6.9) is the component field expansion of the
following superfield Lagrangian
\beqa \L^{conf} &=& \frac{2}{3 \cdot (4\pi)^2} \int d^4 \theta
W^2_{\a\b\gamma}
\frac{1}{\Box_0} \left( (3b-b') R^{\dagger} - \frac{\kappa^2}{24} D^2 K
\right) \nonumber \\
&-& \frac{1}{12 \cdot (4\pi)^2} \int d^4 \theta
\left( \bar{\D}^2 - 8 R\right) \left(
G^2_{\a\dota} - 4 R^{\dagger} R \right)
\frac{1}{\Box_0} \left( b' R^{\dagger} -
\frac{\kappa^2}{24} (d_M - 1) D^2 K \right) \nonumber \\
&+& hc \eeqa 
We conclude, then, that the one-loop three-point graphs shown in
Figure~4a, b and c give rise to the supersymmetric effective
\lag\ (6.9), or, equivalently, to the superfield \lag\ (6.11).
Note that the background field dependent metric
$g_{A A^{\dagger}} - \frac{\kappa^2}{3} G_A G_{A^{\dagger}} $,
which multiplies the $A \bar{A}$-vertices in (4.26), has dropped
out in the effective Lagrangian (6.9).  Hence, the anomaly coefficient
in (6.11) is background field independent.
Using (5.12), it follows that under super-Weyl-\kahler\
transformations
\beqa \delta \L^{conf}
&=& \frac{1}{72 \cdot (4\pi)^2} \int d^2 \theta
\left( 8 (1+3b-b') W^2_{\a\b\gamma} \right.  \nonumber \\
&-& (d_M-1+b')
\left( \bar{\D}^2 - 8 R \right)
\left. \left( G^2_{\a\dota} - 4 R^+ R\right) \right) F
+ h.c. \eeqa 
which is non-vanishing.  Hence, expression (6.11) represents the
contribution to the supersymmetric mixed
super-Weyl-\kahler-Lorentz anomaly from graphs constructed out of
conformal vertices only.

Let us now turn to the contributions from triangle graphs
constructed out of the conformal breaking $A^2$ and $\bar{A}^2$
vertices.  The only possible graphs are shown in  Figure~4d.
These graphs contribute the following non-local term to the
effective \lag\
\beqa \L^{nonconf} &=& \frac{1}{12} \; \frac{1}{18} \;
\frac{| \langle q \rangle |^2}{(4 \pi)^2} \;
\left\{ \left( \R - \frac{{\kappa}^2}{2}
(4 K \left|_{\theta^2
\bar{\theta}^2} \right. + \Box_0 K \left|) \right.
\right)^2 \right. \nonumber \\
&& \left. \times \frac{1}{\Box_0} \left( \R + {\kappa}^2
(4K \left|_{\theta^2
\bar{\theta}^2}\right. + \Box_0 K \left| \right. ) \right)
\right\} \eeqa 
where $q$ is given by
\beq q = \frac{    G_{A A } - \frac{\kappa^2}{3} G_{A
} G_{A}  }{ g_{A A^{\dagger}} - \frac{\kappa^2}{3}
G_{A} G_{A^{\dagger}} } \eeq 
and is the factor multiplying the $A^2$ vertex, as can be seen
from (4.26).  Factor $q$ is background field dependent, but
invariant under K\"{a}hler transformations
${\kappa}^2 K \ra {\kappa}^2 K + F + \bar{F} $, since it is
given in terms of the K\"{a}hler invariant combination
$G = K + ln W + ln \bar{W} $.
Note that it is not $q$, but the constant vacuum expectation
value $\langle q \rangle$ that appears in (6.13).
This occurs, because when calculating the
anomalous contributions from the graphs discussed
in this paper, one actually first expands the background
field dependent metrics in Lagrangians (4.26) and (4.33)
around constant
vacuum expectation values $\langle A \rangle$,
$\langle \bar{A} \rangle$.
The $( \R - \frac{{\kappa}^2}{2} (4 K |_{\theta^2
\bar{\theta}^2} + \Box_0 K |)$-legs in (6.13)
arise through the coupling to the $A^2$ and ${\bar A}^2$
vertices, whereas the
$\R + {\kappa}^2 (4K |_{\theta^2
\bar{\theta}^2} + \Box_0 K |) $ -leg arises through the
coupling to the ${\bar A} A$ vertex.
Lagrangian (6.13) is obviously not invariant
under super-Weyl-K\"{a}hler transformations
and, hence, anomalous.
The component expression (6.13) is
contained in the following superfield \lag\
\beqa \L^{nonconf} &=& \frac{4 | \langle q \rangle |^2}{(4\pi)^2}
\int d^4 \theta {\bar{D}}^2 \left\{ \left(
R^{\dagger} + \frac{{\kappa}^2}{24} D^2 K \right)
\left(R + \frac{{\kappa}^2}{24}
\bar{D}^2 K \right) \right\} \nonumber \\
&& \times \frac{1}{\Box_0} \left( R^{\dagger}  - \frac{{\kappa}^2}{12}
D^2 K \right) + h.c. \eeqa 
Consequently, the total contribution to the mixed
super-Weyl-\kahler-Lorentz anomaly is given by the sum of (6.11)
and (6.15)
\beq \L^{\rm total} = \L^{conf} + \L^{nonconf} \eeq 
We point out that expression (6.16)  could also have been
obtained by direct three-point supergraph calculations using the
prepotential formalism \cite{19,8}.

\lag\ (6.16) represents the mixed super-Weyl-\kahler-Lorentz
anomaly in conventional superspace.  We would now like to
transform it over into \kahler\ superspace.  Following the
manifestly supersymmetric procedure described at the end of
section~5, we now apply the transformations (5.14) to the
anomalous \lag\ (6.16).  The result is
\beqa \L^{total} &=& \frac{1}{12(4\pi)^2} \int d^4 \theta \left( 8
(3b-b')
W^2_{\a\b\gamma} - b' \left( \bar{\D}^2 - 8R \right) \left(
G^2_{\a\dota} - 4 R^{\dagger} R \right) \right) \frac{1}{\Box_0}
R^{\dagger} \nonumber \\
&& + \frac{4 | \langle q \rangle
|^2}{(4\pi)^2} \int d^4 \theta {\bar{D}}^2 \left(
R^{\dagger} R \right) \frac{1}{\Box_0} R^{\dagger} \nonumber \\
&& - \frac{1}{24} \frac{{\kappa}^2}{(4\pi)^2} \int
d^4 \theta \left( \frac{2}{3} (1+3b-b')
W^2_{\a\b\gamma} \right. \nonumber \\
&&- \frac{(d_M-1+b')}{12}  \left.
\left( \bar{\D}^2 - 8R
\right) \left( G^2_{\a\dota} - 4 R^{\dagger} R\right)\right)
\frac{1}{\Box_0} D^2 K \nonumber \\
&& - \frac{| \langle q \rangle |^2}{2}
\frac{{\kappa}^2}{(4\pi)^2} \int d^4 \theta {\bar D}^2
\left( R^{\dagger} R\right) \frac{1}{\Box_0} D^2 K + h.c. \eeqa 
where the \lag\ (6.17) is given in terms of \kahler\ superspace
superfields $W_{\a\b\gamma}$, $G_{\a\dota}$, $R$, $K$.
The first two terms in (6.17) are invariant under K\"{a}hler
transformations, as can be readily seen from (2.6).  Hence, they
can be discarded.  The K\"{a}hler anomaly is entirely given
by the last two terms
\beqa \L^{anom} &=&
- \frac{1}{24} \frac{{\kappa}^2}{(4\pi)^2} \int
d^4 \theta \left( \frac{2}{3} (1+3b-b')
W^2_{\a\b\gamma} \right. \nonumber  \\
&&- \frac{(d_M -1+b')}{12}  \left.
\left( \bar{\D}^2 - 8R
\right) \left( G^2_{\a\dota} - 4 R^{\dagger} R\right)\right)
\frac{1}{\Box_0} D^2 K \nonumber \\
&& - \frac{| \langle q \rangle |^2}{2}
\frac{{\kappa}^2}{(4\pi)^2} \int d^4 \theta {\bar D}^2
\left( R^{\dagger} R\right) \frac{1}{\Box_0} D^2 K + h.c. \eeqa 
These terms are obviously
not invariant under K\"{a}hler transformations.
Comparing the term proportional to $W^2_{\a\b\gamma}$
in (6.18)
with the minimal expression (3.7) shows that both agree!
Lagrangian (6.18) thus qualifies to be called
the supersymmetric \kahler-Lorentz anomaly in \kahler\
superspace.
In component fields, the anomalous
Lagrangian (6.16) reads
\beqa \L^{anom} &=& - \frac{{\kappa}^2}{24 \cdot (4 \pi)^2}
\R_{mna}\,^b \; \tilde{\R}^{mn\;\;a} \!\!\!_b \frac{1}{\Box_0}
\partial^p a_p \nonumber \\
&& + \frac{{\kappa}^2}{12 \cdot 24 \cdot (4\pi)^2}
\left( (d_M +3b) C_{mnpq} C^{mnpq}
- (d_M - 1+b') GB \right) \frac{1}{\Box_0} \nonumber \\
&& \times \left( \left[ 4 K \left|_{\theta^2
\bar{\theta}^2} \right. + \Box_0 K \left| \right. \right]
\right) \nonumber  \\
&& +\frac{1}{144} \; \frac{| \langle
q \rangle |^2 {\kappa}^2}{(4 \pi)^2} \;\;
{\R}^2 \frac{1}{\Box_0} (4K \left|_{\theta^2
\bar{\theta}^2}\right. + \Box_0 K \left| \right. ) + \ldots
\eeqa 
where $\ldots$ refers to terms containing fermions that we have not
explicitly computed.
Let us emphasize again that the coefficient $\langle q \rangle$ of
the ${\cal R}^2 K |$-piece in (6.19) is background field dependant.
This is to be contrasted with the coefficients of the $C_{mnpq}^2
K |$ - and $GB K |$-pieces in (6.19), which are background field
independent.

Finally, we point out that the anomalies involving the
sigma-model Christoffel connection can now easily be computed.
This is done by replacing , in the conventional superspace
calculation, ${\kappa}^2 K$ by ${\kappa}^2 K - 3 ln g$
where $g$ is the K\"{a}hler metric for the chiral multiplet
in the loop.  Also, one can sum over all matter sectors,
denoted by $\sum_S$, and over all matter multiplets, $n_S$, in
a given sector.  Furthermore, it is also not difficult to
extend our results to include gauge vector multiplets
running around the internal loop.  The only changes are that
the $b$ and $b'$ coefficients are now to be evaluated for gauge
vector multiplets, and the unknown coefficient $d_M$, which
ocurred for chiral multiplets, here does not appear.  Putting
all of this together we find, after some manipulation, that
\beqa \L^{anom} &=&
- \frac{1}{36} \frac{{\kappa}^2}{(4\pi)^2} \int
d^4 \theta \left( 3b W^2_{\a\b\gamma}  - \frac{b'}{8}
(8 W^2_{\a\b\gamma} + \left( \bar{\D}^2 - 8R
\right) \left( G^2_{\a\dota} - 4 R^{\dagger} R\right)\right)
\nonumber \\
&& \times \frac{1}{\Box_0} D^2 K \nonumber \\
&& + \frac{1}{24} \frac{1}{(4\pi)^2} \int
d^4 \theta W^2_{\a\b\gamma} \frac{1}{\Box_0} D^2 [
\sum_S (2 ln det g_S - \frac{2}{3} n_S {\kappa}^2 K)] \nonumber \\
&& - \frac{1}{8 \cdot 24} \frac{1}{(4\pi)^2} \int
d^4 \theta \left( \bar{\D}^2 - 8R
\right) \left( G^2_{\a\dota} - 4 R^{\dagger} R\right) \nonumber \\
&& \times \frac{1}{\Box_0} D^2 [ \sum_S (2 ln det g_S
- \frac{2}{3} n_S (d_M - 1) {\kappa}^2 K ) ] \nonumber \\
&& + \frac{| \langle q \rangle |^2}{2}
\frac{1}{(4\pi)^2} \int d^4 \theta {\bar D}^2
\left( R^{\dagger} R\right) \frac{1}{\Box_0} D^2
[ \sum_S (2 ln det g_S - n_S {\kappa}^2 K)] \nonumber \\
&& + h.c. \eeqa 
where
\beqa b &=& \frac{1}{15} (N_S + 3N_F + 12N_V)  \nonumber \\
b' &=& \frac{1}{15} (N_S + \frac{11}{2} N_F + 62N_V)  \eeqa
are the trace anomaly coefficients summed over all matter
and gauge vector supermultiplets.  Note from (6.10) that
the term proportional to $b$ contains in components a
pure $(C_{mnpq})^2$-term, whereas the term proportional
to $b'$ is the superfield expression containing the
Gauss-Bonnet topological density.  In component
fields, the anomalous Lagrangian (6.20) reads
\beqa \L^{anom} &=& - \frac{{\kappa}^2}{36 \cdot (4 \pi)^2}
(3b-b'+N_S)
\R_{mna}\,^b \; \tilde{\R}^{mn\;\;a} \!\!\!_b \frac{1}{\Box_0}
\partial^p a_p \nonumber \\
&& + \frac{{\kappa}^2}{12 \cdot 24 \cdot (4\pi)^2}
\left[ (N_S d_M +3b) C_{mnpq} C^{mnpq}
- (N_S d_M - N_S + b') GB \right] \frac{1}{\Box_0} \nonumber \\
&& \times \left( \left[ 4 K \left|_{\theta^2
\bar{\theta}^2} \right. + \Box_0 K \left| \right. \right]
\right) \nonumber  \\
&& +\frac{N_S}{144} \; \frac{| \langle
q \rangle |^2 {\kappa}^2}{(4 \pi)^2} \;\;
{\R}^2 \frac{1}{\Box_0} (4K \left|_{\theta^2
\bar{\theta}^2}\right. + \Box_0 K \left| \right. ) + \ldots
\eeqa 
where, for simplicity,
we have only displayed the terms proportional to $K$.
Note from (6.21) that $3b-b'+N_S = \frac{3}{2} (N_S
- N_V)$, which yields the correct coefficient in (6.22)
multiplying the term proportional to $\R {\tilde{\R}}$.

We close this section with a brief comment on the relationship
between the mixed supersymmetric
K\"{a}hler-Lorentz anomaly, given by (6.22), and the moduli
dependent threshold corrections to gravitational couplings
\cite{13} in $Z_N$ orbifolds.  These corrections were computed
for the gravitational coupling constant of
$C_{mnpq}^2$.  The coefficient
of these corrections was found to be non vanishing
for $Z_N$ orbifolds
with $N=2$ spacetime sectors and to be proportional to the trace
anomaly coefficients for the different fields coupled to
gravity.  This phenomena can now be somewhat understood in the
field theory framework as follows.  The coefficient of the
$C_{mnpq}^2 K|$-term in (6.22) has, as shown above, its origin
in the coefficients of the two distinctive terms (6.3)
and (6.5).  The coefficient (6.3) is determined by the
chiral anomaly (6.1) computed in the conventional
superspace formulation, whereas the coefficient in (6.5)
is given by the trace anomaly (6.6).  By comparing against
the results of the
explicit string loop calculation of \cite{13} it then follows
that, in $Z_N$ orbifolds with $N=2$ space-time sectors, only
a certain amount of the $C_{mnpq}^2 K|$-term is removed by the
Green-Schwarz mechanism \cite{50}.  This amount is given
by the coefficient of (6.3), yielding a left-over proportional
to (6.6), that is to the trace anomaly coefficients for the different
fields coupled to gravity.  For $Z_N$ orbifolds with no $N=2$
spacetime sectors, the entire amount given in (6.16) must be
removed by the Green-Schwarz mechanism.   This will be discussed
in detail in another publication \cite{100}.

\section{Conclusion}

\hspace{.3in} We have shown how to calculate the matter
contributions
to the
mixed K\"{a}hler-gauge
and mixed K\"{a}hler-Lorentz anomalies
in a manifestly supersymmetric
way.  Even though we have restricted our analysis to
matter loops only, the contributions from
Yang-Mills and gravitational fields running in the loop
can, in principle, also
be computed along the same lines.
All these contributions can be obtained
by first performing calculations in
the conventional formulation of supergravity-matter
systems.  In this formulation, there exists a prepotential superfield
formalism which enables one to separate quantum degrees of freedom
from background fields in a manifestly supersymmetric way.
Calculations in this superfield formalism lead, by construction, to
manifestly supersymmetric results.  Such superfield
calculations of matter loops
should agree
with the component field calculations presented above.
All these results can be transformed over to the K\"{a}hler
superspace formulation of the theory.  In the latter formulation,
the tree-level gravitational kinetic energy is appropriately
Einstein normalised.  The K\"{a}hler anomaly in K\"{a}hler
superspace is obtained from the one in the conventional
formulation by specific superfield rescalings of the underlying
torsion constraints \cite{6}.  Since
these redefinitions
are defined at the superfield level, the resulting expression
for the K\"{a}hler anomaly in K\"{a}hler superspace is
manifestly supersymmetric.

We would like to emphasize the following.
First of all, we have shown that the
procedure described in this paper leads
to the well known result for the mixed K\"{a}hler-gauge anomaly
in K\"{a}hler superspace \cite{7,9,10}, which is responsible for
non-holomorphic threshold corrections to the
running of gauge couplings in string
effective  field theories \cite{7,9,10,11,12,13,40}.
Secondly, the mixed K\"{a}hler-Lorentz anomaly
contains, in general, a term proportional to the square of the
curvature tensor, ${\cal {R}}^2$.  Its coefficient, however,
comes out to be background field dependent, in contrast
to both the gauge case and to the terms in the mixed
K\"{a}hler-Lorentz anomaly proportional to the square
of the Weyl tensor, $C_{mnpq}^2$,
and to the Gauss-Bonnet combination.
As shown in \cite{14,15},
a term in the  K\"{a}hler anomaly proportional
to ${\cal R}^2$ can drive inflation in the early universe
and, hence, it can
have very interesting
cosmological consequences.  Finally, some field theoretical
understanding
of the moduli dependent threshold corrections to gravitational
couplings in $Z_N$ orbifolds \cite{13} can be gained
by comparing our results to the ones obtained in \cite{13}
by an explicit string loop calculation.  These threshhold
corrections, computed for
the gravitational coupling of the
$C_{mnpq}^2$-term,
turn out to be proportional to the trace
anomaly coefficients of the different fields coupled to
gravity.  From the field theory point of view this
can be understood by noticing that there are two
distinct terms which contribute to the K\"{a}hler anomaly,
as described above.  One of them comes with a coefficient
proportional to the trace anomaly whereas the other one
comes with a coefficient proportional to the chiral anomaly
in conventional superspace.  It is this latter contribution
which gets removed by the Green-Schwarz mechanism \cite{50},
yielding threshold corrections proportional to the trace anomaly
coefficients of the various fields coupled to gravity.

\section*{Figure Captions}

\noindent Figure 1a: Fermionic contribution to the
mixed K\"{a}hler-gauge anomaly.

\noindent Figure 1b: Fermionic
contribution to the mixed K\"{a}hler-Lorentz
anomaly.

\noindent Figure 2a: Fermionic
contribution to the mixed super-Weyl-K\"{a}hler-
gauge anomaly.

\noindent Figure 2b: Scalar
contribution to the mixed super-Weyl-K\"{a}hler-
gauge anomaly.

\noindent Figure 2c: Fermionic
and
scalar contributions to the mixed super-Weyl-K\"{a}hler-
gauge anomaly.

\noindent Figure 3: The gauge two-point function.

\noindent Figure 4a: Fermionic
contribution to the mixed super-Weyl-K\"{a}hler-
Lorentz anomaly.

\noindent Figure 4b: Scalar
contribution to the mixed super-Weyl-K\"{a}hler-
Lorentz anomaly constructed from conformal vertices, only.

\noindent Figure 4c: Fermionic
and scalar contributions to the mixed
super-Weyl-Lorentz anomaly constructed from conformal vertices,
only.

\noindent Figure 4d: Graphs contributing
to the mixed super-Weyl-K\"{a}hler-Lorentz
anomaly constructed from non-conformal vertices.

\end{document}